\documentclass[a4paper,11pt]{article}
\usepackage{pos}
\usepackage[backend=bibtex,url=false,doi=false,hyperref,style=numeric,sorting=none,bibencoding=utf8,maxnames=2,uniquename=init,giveninits=true]{biblatex}
\usepackage{amsmath}
\usepackage{braket}
\usepackage{physics}
\usepackage{booktabs}
\usepackage{bbm} 
\usepackage{bm} 
\usepackage{wrapfig}
\newlength{\ftsize}

\title{A qubit regularization of asymptotic freedom without fine-tuning }

\author[a,b]{Sandip Maiti}
\author[a,b]{Debasish Banerjee}
\author[c]{Shailesh Chandrasekharan}
\author*[d]{Marina Krsti\'{c} Marinkovi\'{c}}

\affiliation[a]{Saha Institute of Nuclear Physics, 1/AF Bidhannagar, Kolkata 700064, India}
\affiliation[b]{Homi Bhabha National Institute, Training School Complex, Anushaktinagar, Mumbai 400094, India}
\affiliation[c]{Duke, University, Department of Physics, Box 90305, Durham, NC 27708, USA }
\affiliation[d]{Institut f\"ur Theoretische Physik, ETH Z\"urich,  Wolfgang-Pauli-Str. 27, 8093 Z\"urich, Switzerland }

\emailAdd{sandip.maiti@saha.ac.in}
\emailAdd{debasish.banerjee@saha.ac.in}
\emailAdd{sch27@duke.edu}
\emailAdd{marinama@ethz.ch}

\date{October 2023}
\abstract{

Other than the commonly used Wilson’s regularization of quantum field theories (QFTs), there is a growing interest in regularizations that explore lattice models with a strictly finite local Hilbert space, in anticipation of the upcoming era of quantum simulations of QFTs. A notable example is Euclidean qubit regularization, which provides a natural way to recover continuum QFTs that emerge via infrared fixed points of lattice theories.  
Can such regularizations also capture the physics of ultraviolet fixed points?   
We present a novel regularization of the asymptotically free massive continuum QFT that emerges at the 
Berezenski-Kosterlitz-Thouless (BKT)
transition through a hard core loop-gas model, discussing the advantages this model provides compared to traditional regularizations. In particular, we demonstrate that without the need for fine-tuning, it can reproduce the universal step-scaling function of the classical lattice XY model in the massive phase as we approach the phase transition.

}

\FullConference{The 40th International Symposium on Lattice Field Theory (Lattice 2023)\\
July 31st - August 4th, 2023\\
Fermi National Accelerator Laboratory\\}

\begin{document}

\maketitle

\section{Introduction}
\label{sec:intro}
\label{subsec:o2}
In Wilson’s non-perturbative regularization  
with the space-time lattice, one constructs a lattice Hamiltonian with a quantum critical point where the long distance lattice physics can be argued to be the desired continuum quantum field theory (QFT) of interest,  
inter alia, due to the Gaussian nature of the regularized theory. Can the same be said for discretizations that are not inherently Gaussian? 
Indeed, universality suggests that there is a lot of freedom in choosing the microscopic lattice model to study a particular QFT, one of which goes under the name qubit regularization \cite{Singh:2019uwd,Bhattacharya:2020gpm,Singh:2019jog}, and is the focus of this work. 
    \begin{figure}[t]
\vspace*{-2.3em}
        \centering
        \includegraphics[width=0.43\textwidth]{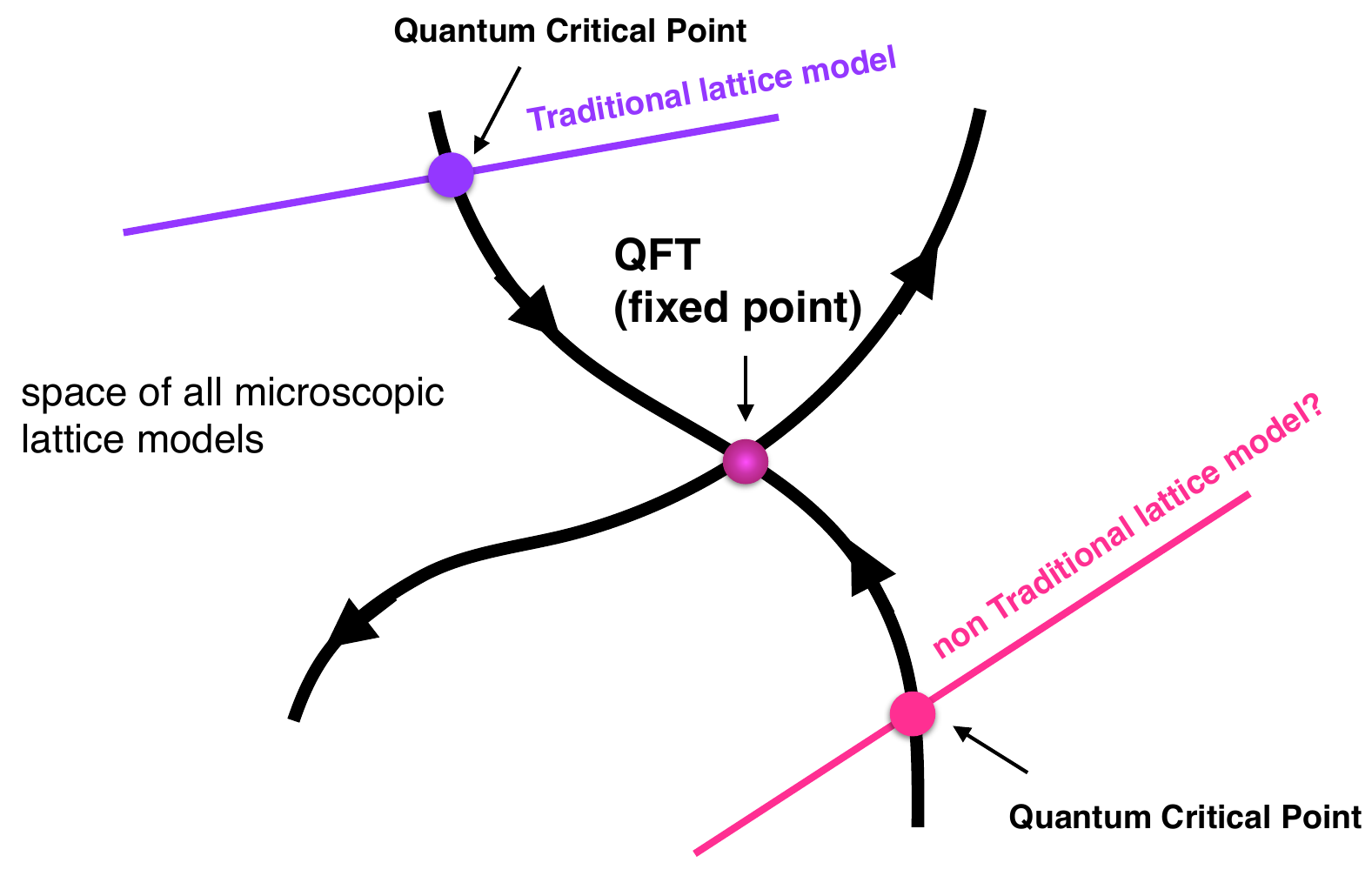}   
\hspace{0.05\textwidth}
     \includegraphics[width=0.5\textwidth]{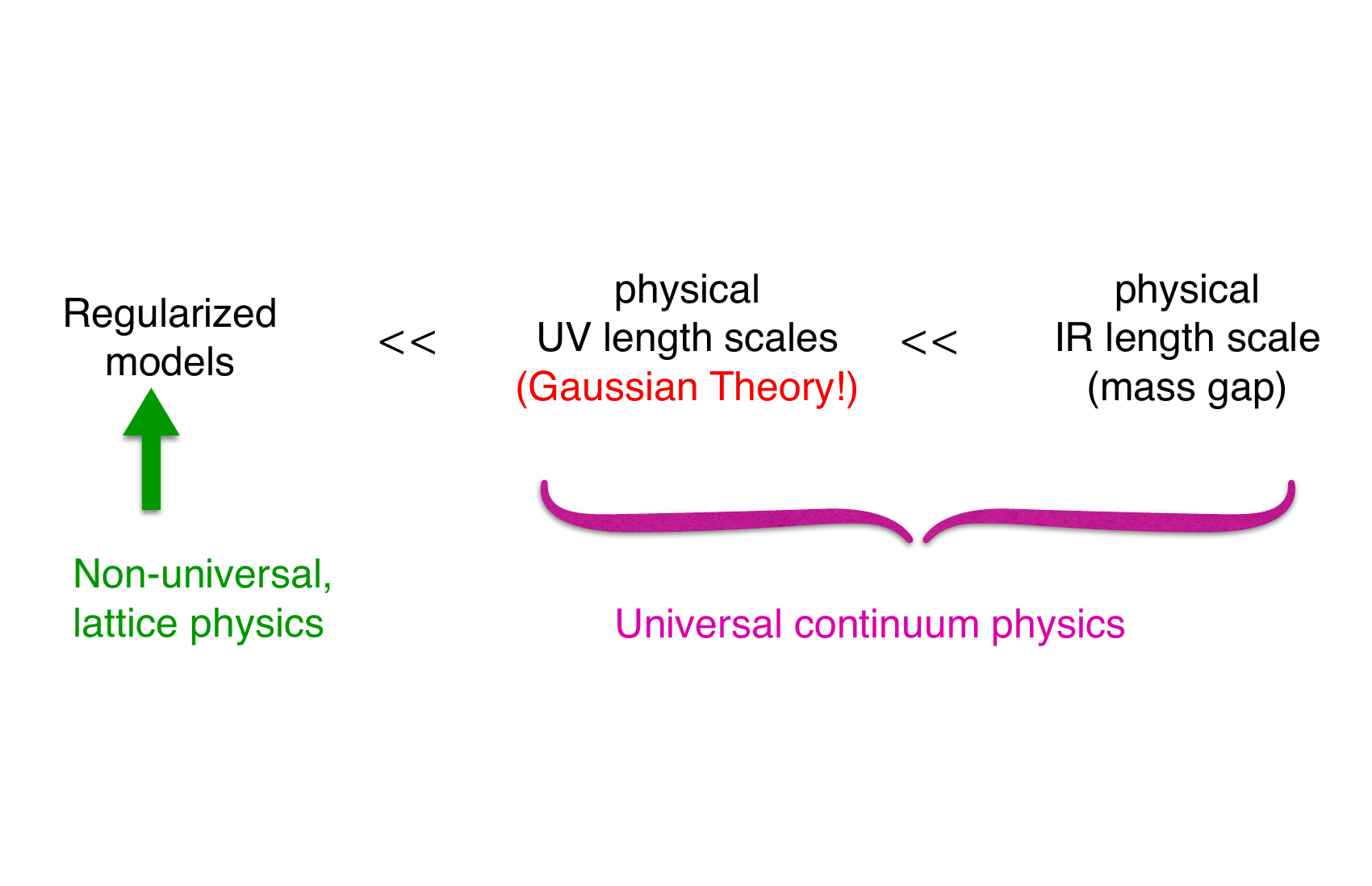}   
        \caption{Wilson's Renormalization Group at work. Different discretizations of the continuum theory lead to the continuum QFT of interest (left). If a discretization is not inherently Gaussian (e.g. qubit discretization), can we nevertheless recover the Gaussian continuum theory (right) by some more complicated RG flow? }
        \label{fig:RG}
\vspace*{-0.5em}
    \end{figure}

In this work we explore if qubit regularization can reproduce the massive physics of a specific two dimensional asymptotically free quantum field theory. 
Concretely, we write the qubit regularized Euclidean theory in terms of a dimer model that can be viewed as a limiting case of a system with two flavors of staggered fermions, originally introduced to study the physics of symmetric mass generation~\cite{Ayyar:2015lrd, Ayyar:2014eua,Maiti:2021wqz}. In this model 
we demonstrate that asymptotic freedom can emerge without fine-tuning due an unconventional qubit regularised model we constructed for the corresponding QFT arising at the BKT fixed point. 
Other than our work in Euclidean formulation of the non-linear O(2) sigma model presented here, 
exponentially small mass gaps were demonstrated in the non-linear O(3) model in  Hamiltonian formulation~\cite{Bhattacharya:2020gpm} as well as  in the $N_f=2$ Schwinger model~\cite{dempsey2023phase}. 

\section{Qubit regularization of QFTs}
\label{sec:qubit}

\begin{wrapfigure}{r}{0.5\textwidth}
        \centering
	\vspace*{-0.5em}
        \includegraphics[width=0.33\textwidth]{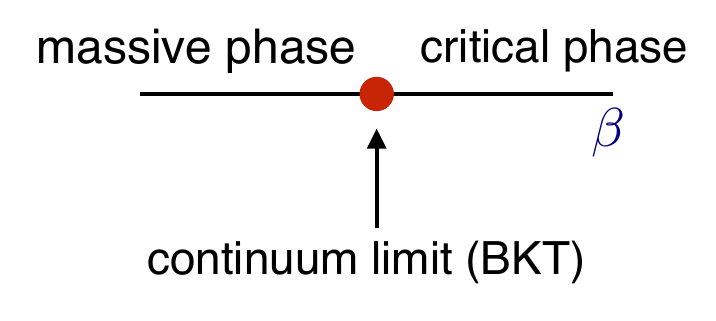}   
        \caption{Traditional (bosonic) regularization of the XY model requires fine tuning to the critical point, where we get the BKT physics in the continuum.}
        \label{fig:BKT}
\end{wrapfigure}
We are interested in the physics of the BKT phase transition, 
and there are many discretizations of the continuum theory to start from, which would eventually lead to the continuum QFT of interest (c.f. left panel of Fig.~\ref{fig:RG}). However, it is not clear whether the Gaussian nature of the UV theory can emerge from these alternative regularizations, especially qubit degrees of freedom, while the same theory then goes on to reproduce the massive physics in the IR? 
For this we need a special type of quantum criticality where the UV and IR length scales emerge simultaneously, distinct from the lattice spacing.
The right panel of Fig.~\ref{fig:RG} illustrates the interplay of these different scales and universality 
starting from a short lattice length scale $a$, 
where we have a variety of regularized models and the non-universal physics depends on factors like the details of qubit regularization.
This is followed by an intermediate length scale, where the continuum UV physics begins to dominate, giving rise to the required Gaussian theory. Eventually, at long length scales, a non-perturbative massive continuum quantum field theory emerges, due to the presence of a marginally relevant coupling in the UV theory.

It is pertinent to discuss some advantages of the lattice formulation of continuum QFTs with a finite local Hilbert space, such as qubit regularization. A notable advantage is  the finiteness of the Hilbert space in a finite lattice volume, making this approach particularly suitable for quantum simulations.
The continuum QFT naturally emerges, upon taking the continuum and 
thermodynamic limits. As outlined in Sec.~\ref{sec:af}, 
in our specific model we are also able to reach the asymptotically free regime in Euclidean space-time without 
fine-tuning. Additionally, we also observe that certain observables in the qubit regularized model show smaller discretization effects compared to those in the traditionally regularized odel.
The specific quantum field theory we are interested in is 
the two-dimensional non-linear O(2) model at the BKT transition, and the two ways to descritize the physics are detailed in the following. 
    \begin{figure}[t]
        \centering
\includegraphics[width=0.60\textwidth]{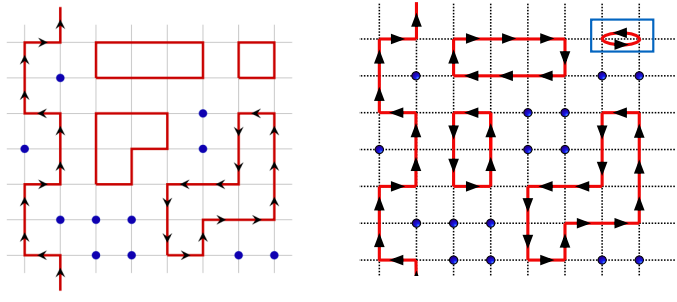} 
\caption{A sample loop configuration in the O(3) model (left panel) and in the O(2) model (right panel), where closed loops on single bonds are now allowed and key to recovering the BKT phase transition without fine-tuning.}
\label{fig:o2o3}
\end{figure}

\section{Non-linear O(2) model: bosonic and fermionic regularization}
\label{sec:o2}
\begin{figure}[t]
\centering
\includegraphics[width=0.7\textwidth]{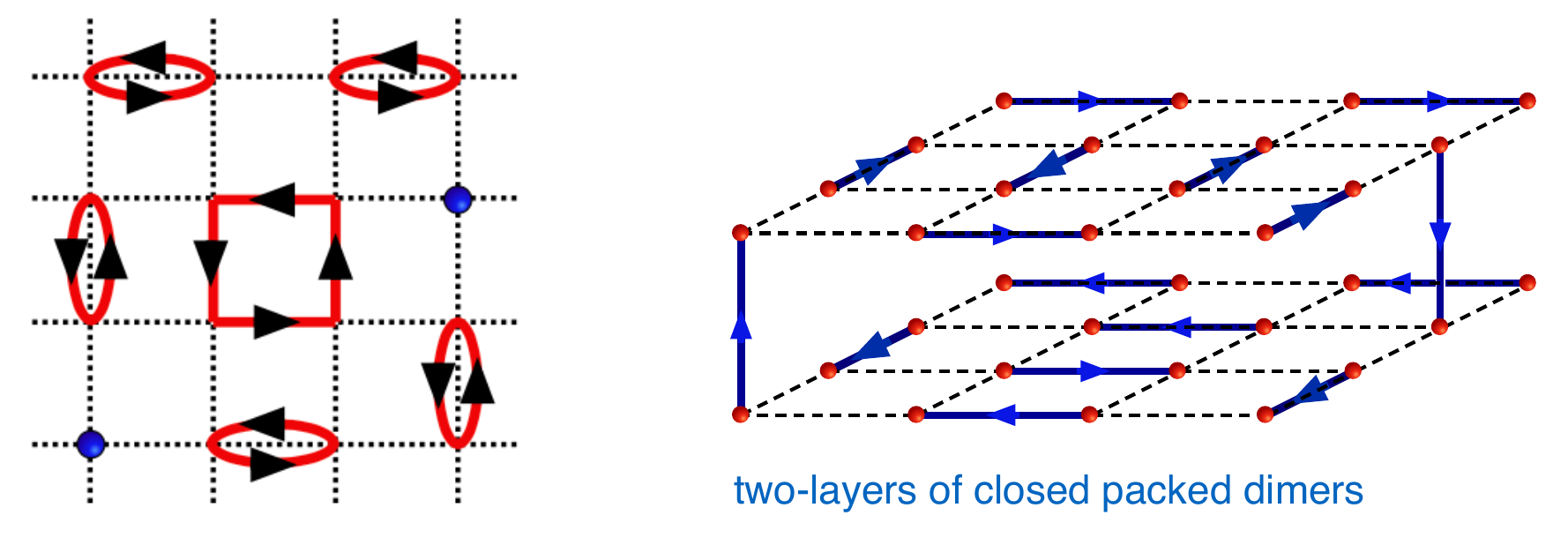}   
\caption{
A sample self-avoiding oriented O(2) loop configuration (left panel) and its representation as a close-packed dimer configuration on two layers (right panel). The t'Hooft vacuum sites, shown as blue circles in the left panel, which have a weight $\lambda$, are mapped to  the interlayer dimers on the right panel, while the  intralayer oriented dimers would form closed oriented loops in the loop representation.
}
\label{fig:Loop}
    \end{figure}
In the traditional discretization of the non-linear O(2) model in two space-time dimensions, the partition function of the lattice model is given by
\begin{align}
Z= \prod_{i}  \int_{0}^{2\pi} d\theta_i ~e^{  {~\beta \sum_{\langle i,j \rangle} cos (\theta_i - \theta_j)}}.
     \label{eq:bosonic}
\end{align}
where the lattice field $0 \leq \theta_i  < 2\pi$ is an angle associated to every space-time lattice site i and $\langle ij \rangle$ denotes the nearest neighbor bonds with sites $i$ and $j$. The lattice field spans an infinite-dimensional Hilbert space of the corresponding one-dimensional quantum model.
It is well known that in this theory an asymptotically free massive continuum quantum field theory arises as one approaches the BKT transition from the massive phase (cf. Fig.~\ref{fig:BKT}). 
 Using high precision Monte Carlo calculations~\cite{Hasenbusch:2005xm,Hasenbusch:2006vd}, the BKT transition has been determined to occur at the fine-tuned coupling of $\beta_c \approx 1.1199(1)$.

This traditional regularization of the non-linear O(2) model can be reformulated in worldline representation and one obtains a partition function in terms of soft core bosons \cite{Banerjee:2010kc}. In contrast 
the qubit regularization amounts to reformulating the partition function as sum over worldline representation with hard-core bosons~\cite{Singh:2019uwd, Bhattacharya:2020gpm}. A qubit regularization of Eq.~(\ref{eq:bosonic}) in Euclidean space-time is represented by the following partition function 
\begin{align}
Z \ = \int [d\bar{\psi} d{\psi}]  \ [d\bar{\chi} d\chi] \ e^{\lambda\sum_i \ \bar{\psi}_i\psi_i \bar{\chi}_i \chi_i} \times e^{ \sum_{\langle ij\rangle}^{~}  \big(\bar{\psi}_i\psi_i\bar{\psi}_j \psi_j + \bar{\chi}_i\chi_i \bar{\chi}_j \chi_j \big)}
     \label{eq:fermionic}
\end{align}
and we will contrast this fermionic model with the commonly used bosonic formulation of Eq.~(\ref{eq:bosonic}). The four Grassmann variables $\psi_i$, $\bar{\psi}_i $, $\chi_i$, and $\bar{\chi}_i $  are defined at each site $i$ of a periodic lattice, coupling $\lambda$ is a weight of the t’Hooft instantons in the system. The configurations contributing to the partition function in Eq.~(\ref{eq:fermionic}) are 
oriented self-avoiding loops (left panel of Fig.~\ref{fig:Loop}), which can also be represented as configurations of closed packed oriented dimers on two layers of square lattices (right panel of Fig.~\ref{fig:Loop}). The coupling  $\lambda$ can be seen in this representation as the weight of the intralayer dimers.  

The model is known to be critical at $\lambda \rightarrow 0$, where the layers of the dimer model decouple. Using worm algorithms for efficent updates of constrained dimer configurations~\cite{Adams:2003cca} we have simulated Eq.~(\ref{eq:fermionic}) for a variety of  lattice sizes $L$ and interlayer couplings $\lambda$. We  demonstrate~\cite{Maiti:2023kpn} that 
when $\lambda > 0$, an asymptotically free massive QFT emerges, as a relevant perturbation at a decoupled fixed point without fine-tuning. 

\section{Asymptotic freedom via decoupled fix point}
\label{sec:af}

\begin{wrapfigure}{r}{0.5\textwidth}
\centering
\includegraphics[width=0.45\textwidth]{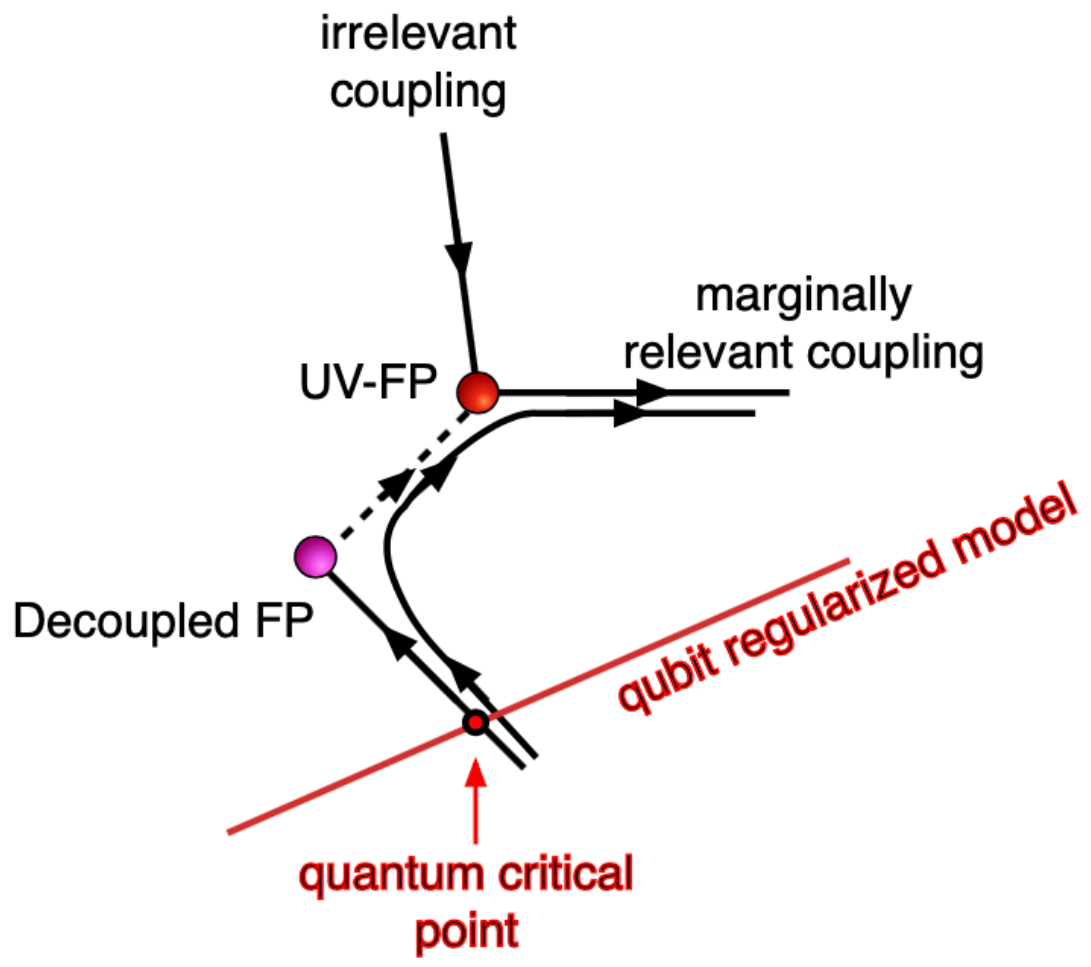}   
\caption{Mechanism of recovering the correct long distance physics once a small non-zero coupling is introduced between the two qubit discretized theories.}
\label{fig:mechanism}
\end{wrapfigure}

The way asymptotic freedom is recovered within qubit regularization is fascinating. One essentially begins with a critical theory which flows to some decoupled fixed point. However, when a small non-zero coupling is introduced the long distance physics becomes massive, but flows through the vicinity of the desired UV-fixed point theory (cf. Fig.~\ref{fig:mechanism}), giving rise to the asymptotically free massive continuum quantum field theory. This RG flow was hidden in the earlier work \cite{Bhattacharya:2020gpm}, but becomes more explicit in our current work where 
we approach the BKT transition using our qubit model.

\begin{figure*}[b]
 \centering
 \includegraphics[width=0.45\textwidth]{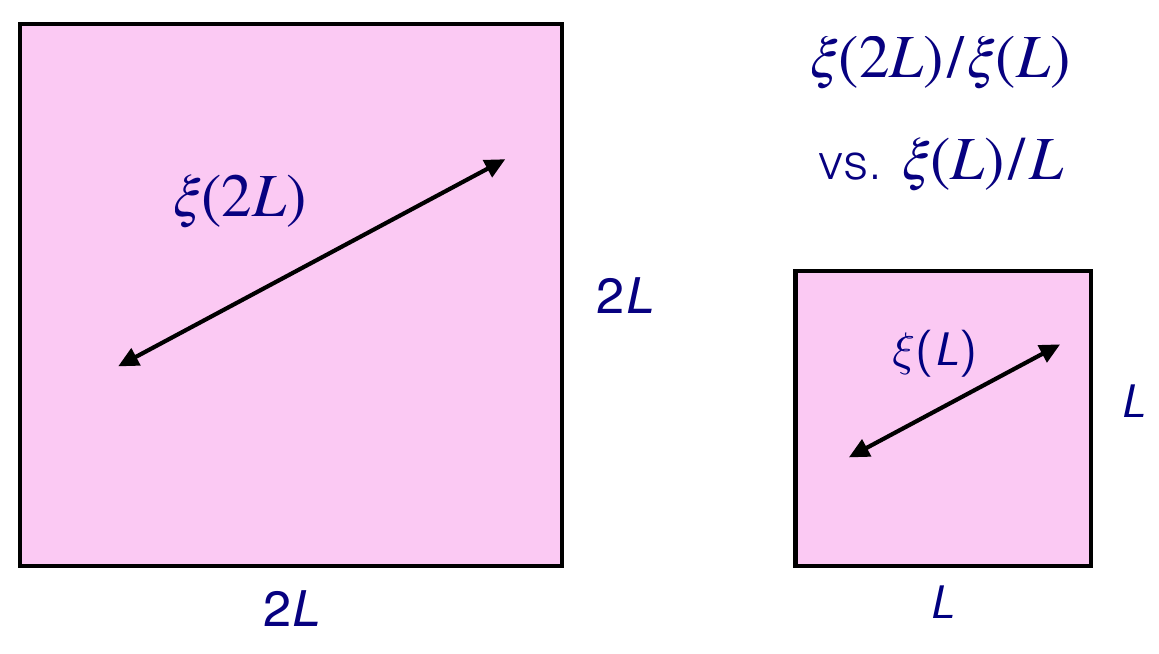}   
\hspace{0.05\textwidth}
 \includegraphics[width=0.30\textwidth]{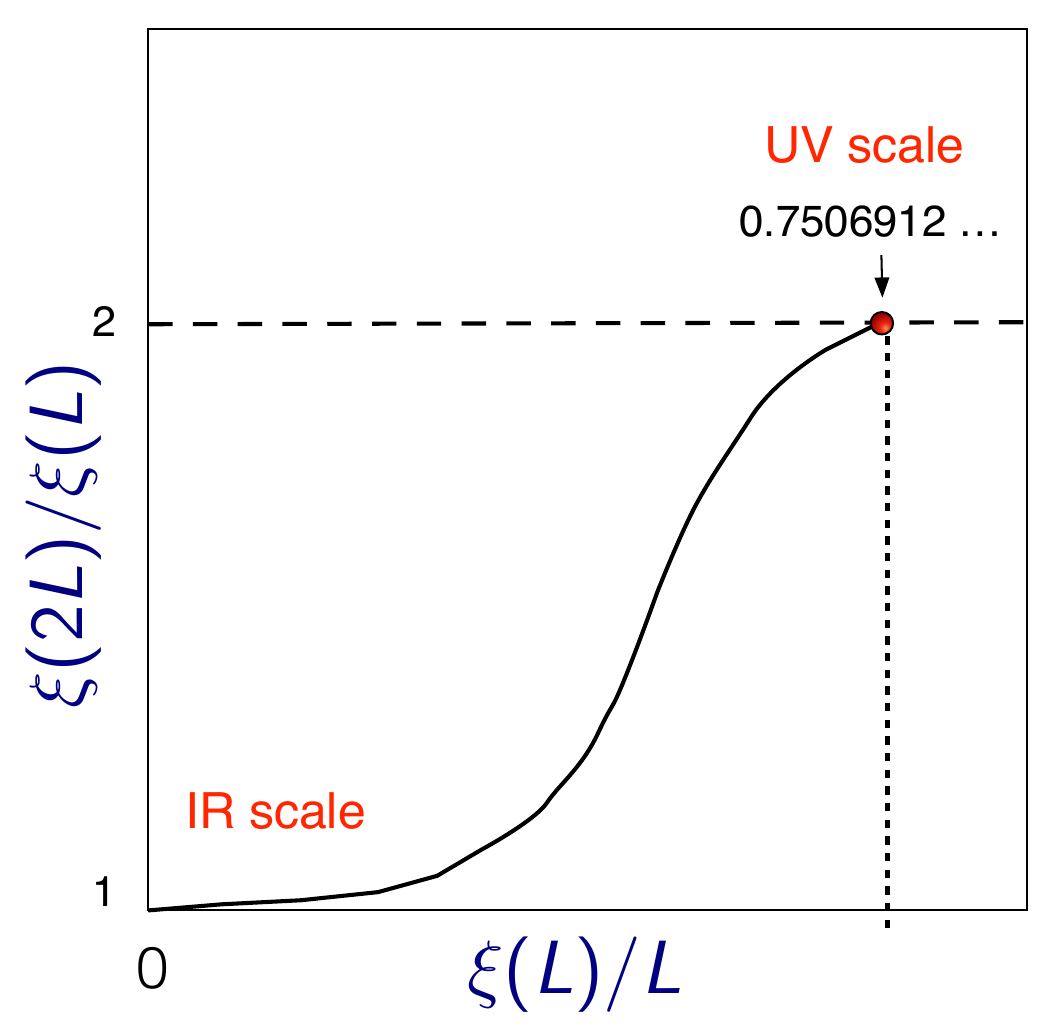} 
 \caption{Sketch of the expected behaviour of the step scaling function defined in Eq.~\ref{eq:scaling} for the asymptotically free theory.}
\label{fig:ScalingSketch}
\end{figure*}

To measure this universal behaviour, we choose the  Caracciolo's step scaling function~\cite{Caracciolo:1994ud} defined as 
\begin{align}
\xi(L)=\frac{\sqrt{\frac{G(0)}{G(2\pi/L)}-1}}{2\:\mathrm{sin}(\pi/L)}
\label{eq:scaling}
\end{align}
where  
$G(p) = \sum_{j\equiv (x,t)} e^{ipx}\langle O^+(t,x) O^- (0,0)  \rangle $
denotes the momentum projected two-point function of either bosonic $O^+_j=e^{i\theta_j}, O^-_j=e^{-i\theta_j}$ or fermionic  $O^+_j=O^-_j=\bar\psi_j\psi_j$ operators. Note that other definitions of the step scaling function, such as e.g. the one introduced in Ref.~\cite{Luscher:1991wu} could have equally been used instead of our chosen defintion. In all these definitions, the step scaling function is expected to follow an universal curve from IR to UV, as sketched in Fig.~\ref{fig:ScalingSketch}. Reproducing this curve with two different models provides evidence that two models are in the same universality class.

\begin{figure*}[t]
\centering
\includegraphics[width=0.54\textwidth]{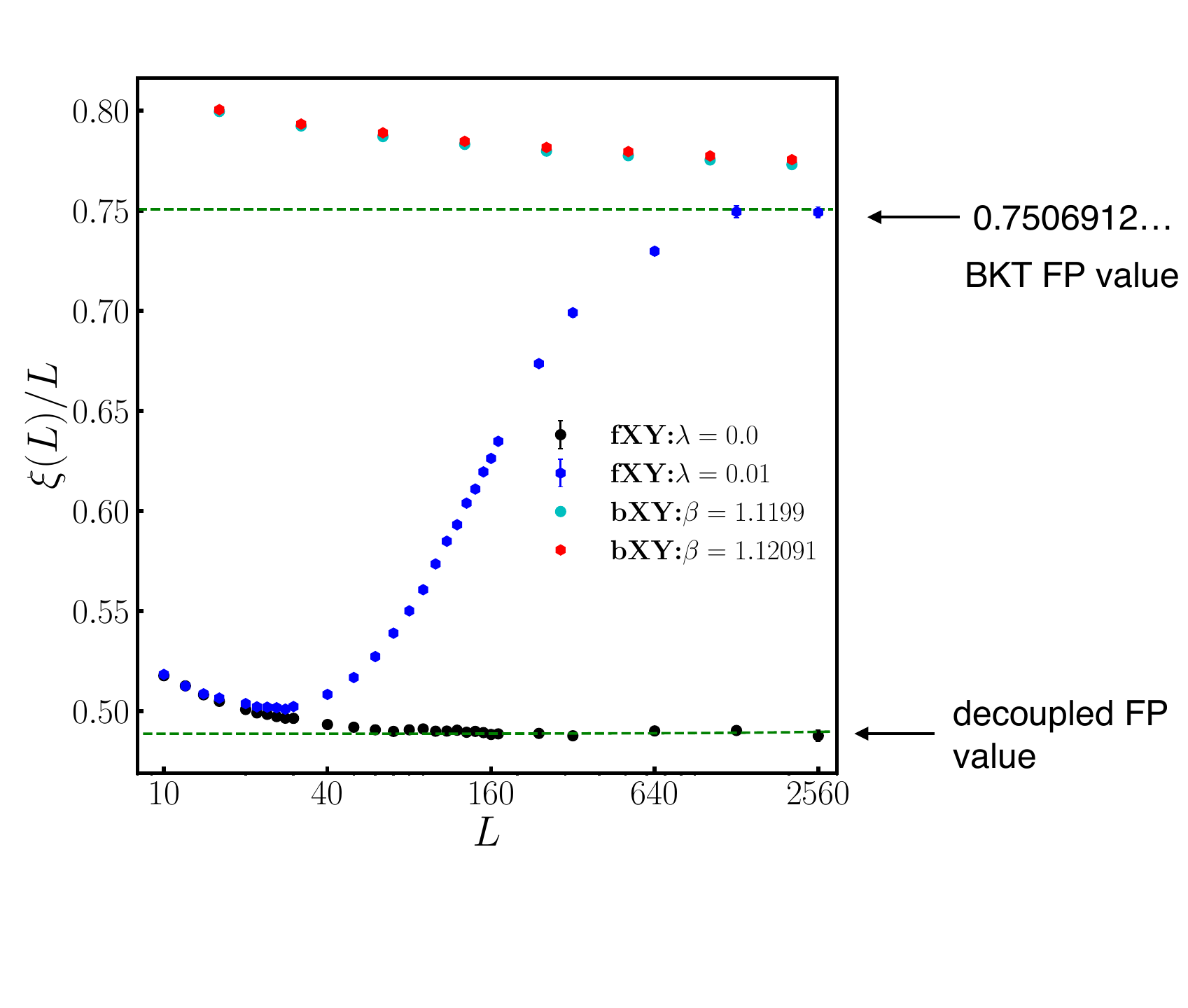}   
\hspace{0.04\textwidth}
\includegraphics[width=0.40\textwidth]{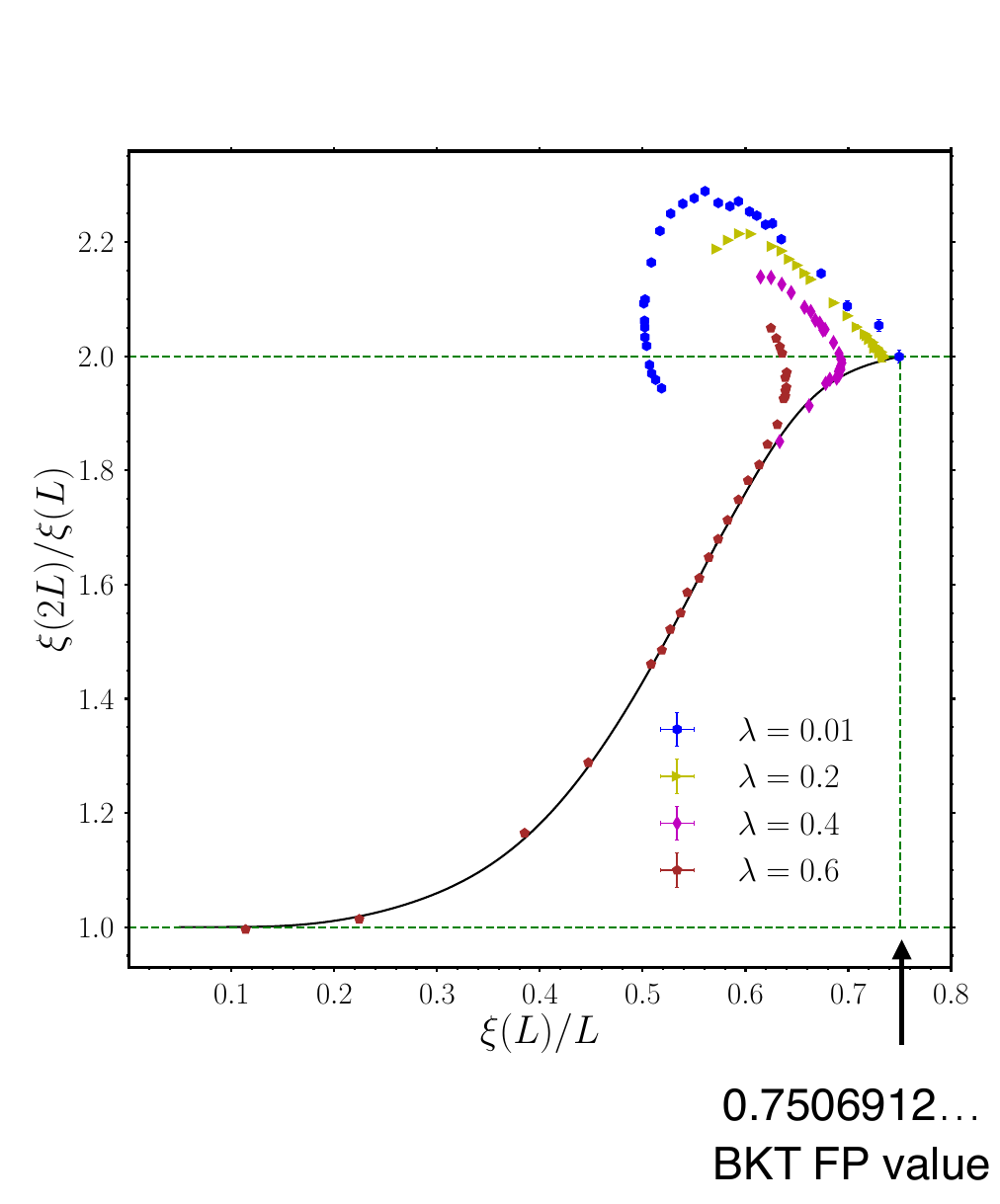}   
\caption{Step scaling function defined in Eq.~\ref{eq:scaling} computed for the bosonic (bXY) and fermionic (fXY) XY model, whose partition functions are given in Eqs.~\ref{eq:bosonic} and ~\ref{eq:fermionic}.
At $\lambda=0.01$ and for large $L$, the data approaches the universal UV prediction $\xi(L)/L = 0.7506912$. 
}
\label{fig:Scaling}
\end{figure*}

The step scaling function computed from the simulations of the bosonic and fermionic XY models are shown in Fig.~\ref{fig:Scaling}. We see that the universal step scaling function of the traditional (bosonic) XY model is reproduced by our fermionic qubit-discretized model, without fine-tuning (for small $\lambda$ instead of only at $\lambda=0$), after which a thermodynamic limit needs to be taken. The matching demonstrates the universal behaviour expected from Wilson's renormalization group.
Moreover, the universal quantities at the UV scale show smaller finite size effects in our model as compared to the traditional XY model (c.f. left panel of Fig.~\ref{fig:Scaling}).
To our knowledge this is the first time a complete universal curve has been computed in an Euclidean box, while the limit $L_{\tau} \rightarrow \infty$ the universal step scaling function for the non-linear O(2) sigma model has been studied in \cite{Balog:2001wv, Balog:2000ra}. 

\begin{figure*}[h]
\centering
\hspace*{0.1\textwidth}
\includegraphics[width=0.7\textwidth]{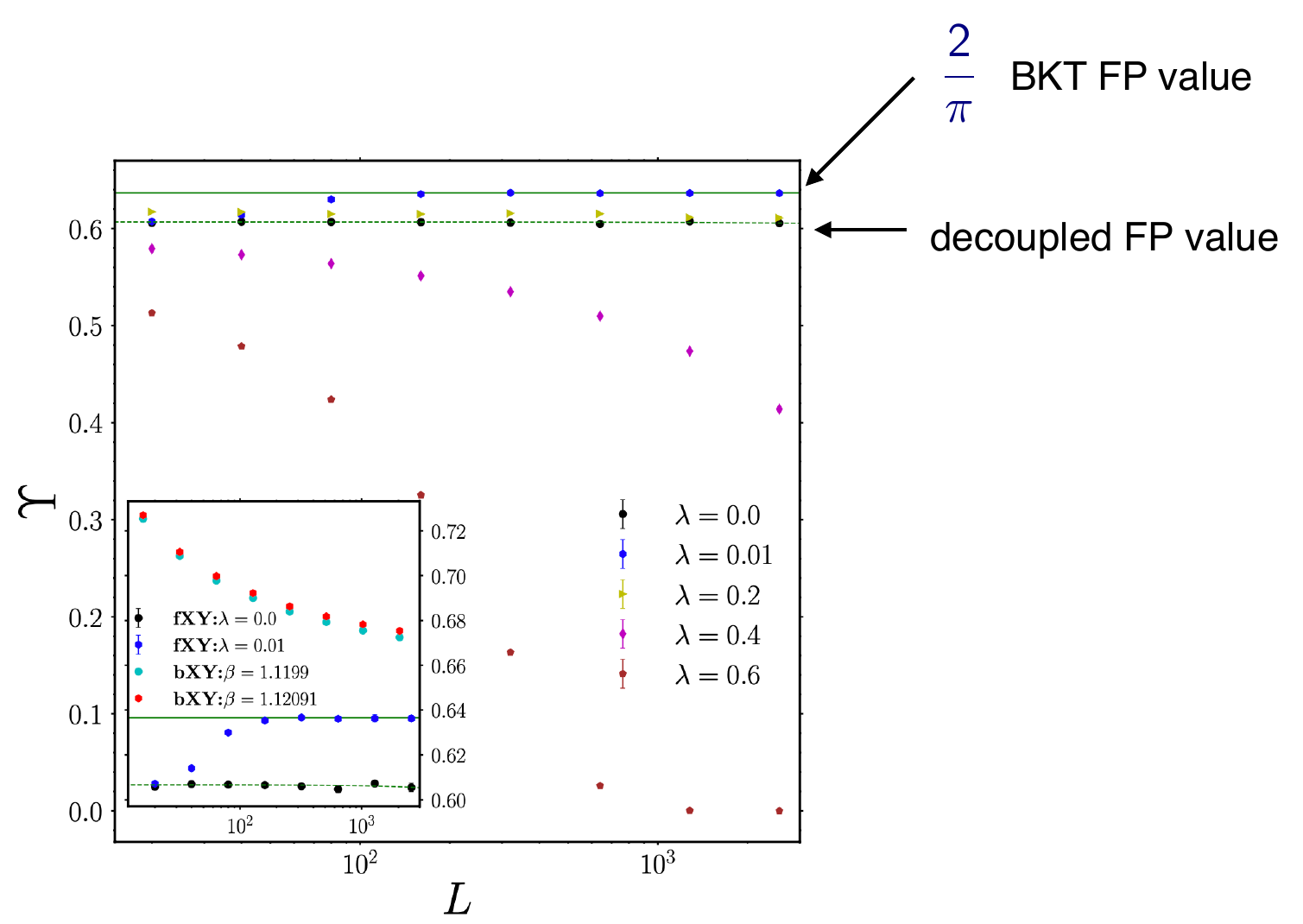}   
\caption{Helicity modulus 
$\Upsilon$ for a variety of lattice sizes $L$ and intralayer coupling $\lambda$. At the UV fixed point the universal value $\Upsilon \approx 2/\pi$ is reached at $\lambda=0.01$ for moderate lattice sizes. Since $\lambda=0.01$  is not actually the critical point, this line will also eventually turn around at very large $L$ values and approach zero.
}
\label{fig:Helicity}
\end{figure*}

\section{Helicity modulus and discretization effects}

In addition to the universal step scaling function discussed in the previous section, we have computed the helicity modulus in both discretizations, $\Upsilon = \langle Q_{\omega}^2\rangle$,  corresponding to the spatial winding number of bosonic worldlines. As expected, we notice the difference in the helicity modulus between $\lambda \rightarrow 0$ and $\lambda= 0$, 
while in the  massive phase $\Upsilon$ must vanish. Similar to the universal step scaling function, in Fig.~\ref{fig:Helicity} we see that we can reach the asymptotically free regime without fine tuning in Euclidean spacetime, albeit with smaller discretization effects: we reach the universal BKT FP value of $2/\pi$ already for lattices $L>10^3$ at $\lambda=0.01$, while a traditional XY model would require lattices $L\gg 10^3$ to achieve comparable discretization effects.

\section{Conclusions}
\label{sec:concl}

We recall that there are many ways to regularize QFTs, and focus on a particular qubit regularization approach, which allows us to recover asymptotically-free massive QFTs using the introduced qubit-degrees of freedom. Unlike the traditional lattice regularization approach there are no guarantees for a connection between the qubit regularized theory to the Gaussian theory, and this needed to be confirmed by numerical experiments. At the example of the two-dimensional non-linear O(2) model (XY model) and its qubit regularization in Euclidean spacetime, we demonstrate that we can reproduce the universal parameters obtained in simulations of the classical bosonic XY model~\cite{Hasenbusch:2005xm}. 

Our interpretation of the underlying mechanism is similar to the one discovered in a Hamiltonian formulation of the two-dimensional nonlinear $O(3)$ sigma model~\cite{Bhattacharya:2020gpm}: at the decoupled quantum critical point, two differently qubit regularized models describe the physics of a critical system containing two decoupled theories. However, when a small non-zero coupling is introduced between the theories, the long distance physics flows towards the desired universal physics of the UV-FP theory (c.f.~Fig~\ref{fig:mechanism}). Our non-traditional qubit regularized model thus gives the continuum physics of the XY model, and the expected physics both in UV and IR is recovered without fine-tuning, at $\lambda$ values close to, but not equal to zero.
Furthermore, we see significantly smaller cut-off effects in universal quantities such as helicity modulus and universal step scaling function, when compared to the same observable computed in the traditional XY model. 
  
\acknowledgments

We are grateful to J. Pinto Barros, S. Bhattacharjee, T. Bhattacharya, H. Liu, S. Pujari, K. Damle, A. Sen, H. Singh and U.-J. Wiese for inspiring discussions. We acknowledge use of the computing clusters at SINP, and the access to Piz Daint at the Swiss National Supercomputing Centre, Switzerland under the ETHZ’s share with the project IDs go24 and eth8. Support from the Google Research Scholar Award in Quantum Computing and the Quantum Center at ETH Zurich is gratefully acknowledged. S.C’s contribution to this work is based on work supported by the U.S. Department of Energy, Office of Science — High Energy Physics Contract KA2401032 (Triad National Security, LLC Contract Grant No. 89233218CNA000001) to Los Alamos National Laboratory. S.C is supported by a Duke subcontract based on this grant. S.C’s work is also supported in part by the U.S. Department of Energy, Office of Science, Nuclear Physics program under Award No. DE-FG02-05ER41368. D.B. acknowledges assistance from SERB Starting Grant No. SRG/2021/000396-C from the DST (Government of India).


 \printbibliography

\end{document}